\documentclass[
aps,
prd,
twocolumn,
superscriptaddress,
nofootinbib,
longbibliography
]{revtex4-2}

\usepackage{amsmath}
\usepackage{amssymb}
\usepackage{bm}

\usepackage{graphicx}
\usepackage{dcolumn}
\usepackage{booktabs}

\usepackage{xspace}
\usepackage{microtype}

\setcitestyle{numbers,sort&compress}

\usepackage[colorlinks=true,
            linkcolor=blue,
            citecolor=blue,
            urlcolor=blue]{hyperref}

\newcommand{\mphi}{\ensuremath{m_\phi}\xspace}

\newcommand{\rhoDM}{\ensuremath{\rho_\mathrm{DM}}\xspace}
\newcommand{\GeV}{\ensuremath{\,\mathrm{GeV}}\xspace}

\newcommand{\Ms}{\ensuremath{M_s}\xspace}
\newcommand{\fs}{\ensuremath{f_s}\xspace}
\newcommand{\fphi}{\ensuremath{f_\phi}\xspace}

\begin{document}

\title{Solar System Probes for Scalar Field Dark Matter}

\author{Edmund Ghampson}
\affiliation{Department of Physics and Astronomy, Vanderbilt University,
Nashville, Tennessee 37235, USA}

\author{Oem Trivedi}
\affiliation{Department of Physics and Astronomy, Vanderbilt University,
Nashville, Tennessee 37235, USA}

\author{Robert J. Scherrer}
\affiliation{Department of Physics and Astronomy, Vanderbilt University,
Nashville, Tennessee 37235, USA}

\author{Alfredo Gurrola}
\affiliation{Department of Physics and Astronomy, Vanderbilt University,
Nashville, Tennessee 37235, USA}

\date{\today}
\begin{abstract}
Scalar field dark matter provides us with a well motivated alternative to conventional particle dark matter, especially when ultralight fields form coherent oscillations or compact self gravitating clumps. Here we develop three complementary Solar System and local Galactic level probes of such models. These probes pertain to ADAF-like flares from scalar clump encounters with Kuiper Belt Objects, atomic clock searches for oscillatory variations of fundamental constants and astrometric microlensing by compact scalar clumps. We derive simple sensitivity estimates and null detection bounds on the scalar clump fraction, clock couplings and compact lens abundance. Our results show that Gaia-like astrometry can probe compact scalar clumps at the percent level near $M_s\gtrsim10^{-2}M_\odot$, while future astrometric and clock experiments can extend the reach to lower masses and weaker couplings.
\end{abstract}

\vspace{2cm}


\maketitle

\section{Introduction}
Multiple studies of astrophysical and cosmological evidence indicate the presence of nonbaryonic dark matter, yet its microscopic nature remains unknown \cite{bertone2005particle, feng2010dark}. The existence of dark matter can be inferred through a remarkably different set of independent observations spanning many decades of length scale. On galactic scales, the anomalously flat rotation curves of spiral galaxies provide perhaps the most direct evidence, with the inferred mass profiles extending far beyond the visible stellar disc \cite{begeman1991extended}. On the scale of clusters, the virial theorem applied to galaxy velocities, X-ray observations of the interstellar gas temperature and luminosity \cite{allen2002cosmological}, and the gravitational lensing \cite{bartelmann2010gravitational} all independently point to a dominant matter component. Moreover, on the scale of cosmology,  accurate measurements of the cosmic microwave background (CMB) anisotropies show that conventional baryonic matter makes up only about $\approx 4.9 \%$ of the universe's total energy budget, but dark matter makes up around $\approx26.8\%$ \cite{aghanim2020planck}. Constraints from Big Bang nucleosynthesis (BBN) and baryon acoustic oscillations independently affirm that the majority of this matter cannot be baryonic, establishing nonbaryonic dark matter as a fundamental component of the standard cosmological model ($\Lambda$CDM) \cite{cooke2018one}.

\indent Among the well-motivated alternatives to conventional particle dark matter are \emph{ultralight scalar fields} with masses in the range $m_\varphi \sim 10^{-22}$--$10^{-10}$\,eV \cite{Hui:2017ltb, marsh2016axion}. Toward the lower end of this range, fuzzy dark matter (FDM) and axion-like particles provide a particularly interesting class of candidates \cite{dentler2022fuzzy}. Their large occupation numbers allow the dark matter to behave as a coherent field on galactic scales, while in suitable regimes the same fields may form compact, self-gravitating configurations such as boson stars \cite{Liebling:2012fv}. Such objects arise when self-gravity is balanced by gradient or quantum pressure, producing long-lived scalar configurations and, depending on the underlying model, oscillating solutions \cite{Liebling:2012fv, liebling2023dynamical}. Cosmological simulations further indicate that FDM haloes can develop central solitonic cores \cite{Schive:2014dra}, whose density profiles and core--halo relations depend on the mass of the underlying scalar field. A population of smaller scalar substructures may also be generated through processes such as tidal stripping of haloes or parametric resonance during the post-inflationary epoch \cite{arvanitaki2020large}. The abundance, mass distribution, compactness, and spatial distribution of these structures therefore provide a direct connection between the microscopic properties of the scalar field and potentially observable astrophysical signatures.

\indent Scalar dark matter can be investigated through a combination of terrestrial experiments, space based measurements, and astrophysical observations, including atomic clocks, resonant detectors and precision astronomical surveys \cite{Ferreira:2021,van2015search,Kennedy:2020bac,
hees2016searching, Wcislo:2018}. Clock networks operating on the Earth's surface, aboard satellites and within GPS constellations have searched for the oscillatory or transient variations of fundamental constants predicted in scalar dark matter models. Although no statistically significant signal has yet been identified, these searches can place stringent upper limits on the couplings between scalar fields and Standard Model parameters \cite{kobayashi2022search, Roberts:2017}. These measurements should therefore be interpreted as null constraints on the relevant couplings rather than as evidence for a dark matter detection.

\indent In this work, we consider three complementary Solar System and local Galactic probes of scalar field dark matter. First, we derive sensitivity estimates for a possible transient channel in which a compact scalar clump, such as a boson star, encounters a Kuiper Belt Object (KBO) and produces an ADAF-like luminosity flare \cite{Liebling:2012fv}. Photometric microlensing by compact dark matter has been studied extensively, but the possibility of luminosity transients associated with the disruption or strong perturbation of KBOs has received comparatively little attention. We therefore provide first order-of-magnitude estimates of the event rate, luminosity, observational reach, and the resulting constraints on the clump fraction $f_{s}(M_{s})$. Second, we present a unified and self contained framework for translating atomic-clock measurements into constraints on oscillatory variations of fundamental constants induced by a light scalar field. Our treatment is based on the theoretical framework of Refs. \cite{hees2016searching, Kennedy:2020bac, mondino2020first, Wcislo:2018} and is written in a form that can be applied directly to future clock-comparison datasets. The underlying theory and several experimental upper limits are already well established \cite{Kennedy:2024,hees2016searching,Wcislo:2018,kobayashi2022search}. Our purpose is to bring these results into a common framework with the other probes considered here and to make the assumptions and sensitivity scalings of the clock channel explicit. Third, we use the standard formalism of astrometric microlensing \cite{Walker:1995, Dominik:2000, Proft:2011}, which has already been applied to Gaia data in several recent studies \cite{Rybicki:2023,mondino2020first,Demir:2025,Verma:2022}, and adapt it to compact scalar configurations, including boson stars with extended density profiles. The three probes are sensitive to different manifestations of the scalar sector. The flare channel responds to compact clumps capable of depositing energy into ordinary matter, atomic clocks probe coherently oscillating field components through their couplings to Standard Model quantities, and astrometric lensing depends only on the gravitational influence of compact scalar structures. Our aim is to clarify what each channel can independently constrain, identify regions of parameter space in which their sensitivities overlap, and provide a framework through which future null results, or a possible candidate signal can be translated into quantitative constraints on scalar dark matter models. 

\section{ADAF-like Flares from Scalar Dark Matter Clumps}
\label{sec:flares}

In an Advection-Dominated Accretion Flow (ADAF), matter falls into a compact object such as a black hole without losing much energy, less than one percent of its energy. \cite{chandra2008adaf}. ADAFs are believed to exist in low-luminosity active galactic nuclei (AGN) \cite{liu2025generalized} and this provides the bridge between dark matter physics and high energy astrophysical transients, offering a way to detect or constrain exotic dark matter through distinctive signatures produced when compact dark matter structures encounter black holes. We first examine the case in which scalar dark matter forms compact clumps (e.g. boson stars) \cite{Liebling:2012fv} of mass \Ms\ that constitute a fraction \fs\ of the local dark matter density \rhoDM. The number density of clumps is given by
\begin{equation}
  n_s = \frac{f_s\,\rho_\mathrm{DM}}{M_s} \;.
\label{eq:ns}
\end{equation}
for a target body of bulk density $\rho_\mathrm{obj}$ in the outer Solar System; one may define an effective interaction radius $R_\mathrm{int}(M_s)$, for instance, of tidal origin,
\begin{equation}
  R_\mathrm{int} \sim \left(\frac{M_s}{\rho_\mathrm{obj}}\right)^{1/3} \;,
\label{eq:Rint}
\end{equation}
with a corresponding geometric cross-section of
$\sigma_\mathrm{int} \sim \pi R_\mathrm{int}^2$.
For a population of $N_\mathrm{obj}$ targets moving with characteristic relative velocity $v$, the total trigger rate is then
\begin{equation}
  \Gamma_\mathrm{trig} \sim n_s\,N_\mathrm{obj}\,\sigma_\mathrm{int}\,v \;.
\label{eq:Gammatrig}
\end{equation}

Parameterizing the radioactive output through the available gravitational interaction energy, we get
\begin{equation}
  E_\mathrm{avail} \sim \frac{G M_s m_\mathrm{obj}}{R_*}
\label{eq:Eavail}
\end{equation}
where $m_{obj}$ is the target mass and $R_{*}$ is the relevant interaction scale, and if a fraction $\eta$ is radiated over a timescale $\tau$, one may estimate the peak luminosity as
\begin{equation}
  L_\mathrm{pk} \sim \eta\,\frac{G M_s m_\mathrm{obj}}{\tau R_*} \;.
\label{eq:Lpk}
\end{equation}
For an instrument with flux limit $F_\mathrm{lim}$, the corresponding detectability radius is
\begin{equation}
  D_\mathrm{det} \sim \left(\frac{L_\mathrm{pk}}{4\pi F_\mathrm{lim}}\right)^{1/2} \;.
\label{eq:Ddet}
\end{equation}

It is useful to establish the estimations in Eqs. (1)–(6) for the Kuiper Belt (bodies in the outer solar system) as the natural target ensemble. The Kuiper belt is known to have roughly 100 $N_{obj} \sim 10^{5} - 10^{6}$ objects larger than roughly 100 km \cite{bernstein2004size}. Stellar kinematics provide strong constraints on the local dark matter distribution in the vicinity of the solar system $\rhoDM \simeq 0.4 \GeV \; cm^{-3}$ \cite{de2021dark}, and the halo velocity dispersion determines the typical relative velocity between a virialized Galactic dark matter clump (thus an invisible gravitational bound pocket of dark matter) and a Kuiper Belt Object (KBO), $v \sim 220\; kms^{-1}$  \cite{necib2019inferred, staudt2024sliding}.\\
A fiducial KBO bulk density $\rho_{obj} \sim 10^{3}$ $kg m^{-3}$ is consistent with the measured densities of large trans-Neptunian bodies \cite{lacerda2007densities}. Substituting these values into Eqs. (1)-(3) one arrives at a trigger rate 

\begin{equation}
    \Gamma_{\rm trig} \simeq 10^{-4} \, \text{yr}^{-1}
    \left( \frac{f_s}{10^{-2}} \right)
    \left( \frac{M_s}{10^{-10} \, M_\odot} \right)^{\!\!-1/3}
    \left( \frac{N_{\rm obj}}{10^5} \right)
    \label{eq:trigger_rate}
\end{equation}
This rate is low but non-negligible over a decade-long observational period. A survey conducted over ten years and monitoring approximately $\sim 10^{5}$ Kuiper Belt Objects (KBOs) accumulates $\mathcal{O}(1)$ events from clump fractions $f_{s}\sim\geq 10^{-2}$ in a mass range $M_{s} \sim 10^{-12} - 10^{-8}M_\odot$. This mass range corresponds with the mass range possessed by the boson stars produced from the ultralight scalar mass \cite{Hui:2017ltb, Liebling:2012fv}. For a typical KBO with radius $R_{*} \sim \mathrm{500}\;\mathrm{km}$ and mass $m_{\mathrm{obj}} \sim 10^{21}\;\mathrm{kg}$, the peak luminosity can be calculated. The natural orbital crossing time $\tau \sim R_{*}/v \sim 10^{3}\mathrm{s}$ gives the disruptive timescale, which is a typical ADAF-class flow in which most of the dissipated energy is advected inward rather than promptly radiated. one finds 
\begin{equation}
    L_{\mathrm{pk}} \simeq 10^{22} \, \mathrm{W}
    \left( \frac{\eta}{10^{-2}} \right)
    \left( \frac{M_s}{10^{-10} \, M_{\odot}} \right)
    \left( \frac{R_*}{500 \, \mathrm{km}} \right)^{\!-1}
\end{equation}\\
This brightness is comparable to the steady X-ray output of a young neutron star. It is within the range of the next-generation wide field time-domain observatories like the Vera C. Rubin Observatory/LSST \cite{ivezic2019lsst}.

Assuming a survey duration T, monitoring $N_{\mathrm{obj}}$ KBOs with detection efficiency $\epsilon_{det}$, finds no events, which is consistent with the flare signatures. Poisson statistics give a 95\% confidence upper bound $\Gamma_{\rm trig} < 3/(N_{\rm obj}\,T\,\epsilon_{\rm det})$, which via Eqs.~(1)-(3) translates into

\begin{align}
    f_s(M_s) &< \frac{3\,M_s}{\rho_{\rm DM}\,v\,\pi R_{\rm int}^2\,N_{\rm obj}\,T\,\epsilon_{\rm det}} \nonumber \\
    &\simeq 5\times10^{-3}
    \left(\frac{M_s}{10^{-10}\,M_\odot}\right)^{1/3}
    \left(\frac{\rho_{\rm obj}}{10^3\,\mathrm{kg\,m}^{-3}}\right)^{2/3} \nonumber \\
    &\phantom{{}\simeq{}}\times
    \left(\frac{N_{\rm obj}}{10^5}\right)^{-1}
    \left(\frac{T}{10\,\mathrm{yr}}\right)^{-1}
    \left(\frac{\epsilon_{\rm det}}{0.5}\right)^{-1}.
    \label{eq:fs_bound}
\end{align}  
This bound is competitive with current microlensing constraints, and, in some mass windows, it is tighter \cite{niikura2017microlensing}. The loss of sensitivity below $M_{s} \simeq 10^{-11}M_\odot$ is due to finite stellar source size and cadence limitations. By direct coupling to test masses, gravitational wave detectors have been proposed as probes of compact dark objects. \cite{arvanitaki2020large}. However, those constraints hold in a different coupling channel and in a complementary mass regime. Eqn. \ref{eq:fs_bound} shows a mild power-law growth with $M_{s}$, heavier clumps becoming progressively harder to constrain for fixed survey parameters because the number density $n_{s} \propto M^{-1}$ drops faster than the cross section $\sigma_{\mathrm{int}} \propto M_{s}^{2/3}$ grows. In practice, the flare channels are optimally sensitive in the mass range $M_{s} \sim 10^{-14}-10^{-8}M_\odot$, corresponding to boson star masses accessible for scalar field masses $m_{\phi} \sim 10^{-17}-10^{-12}\mathrm{eV}$ \cite{Liebling:2012fv}.
\begin{figure*}[t]
    \centering
    \includegraphics[width=0.48\textwidth]{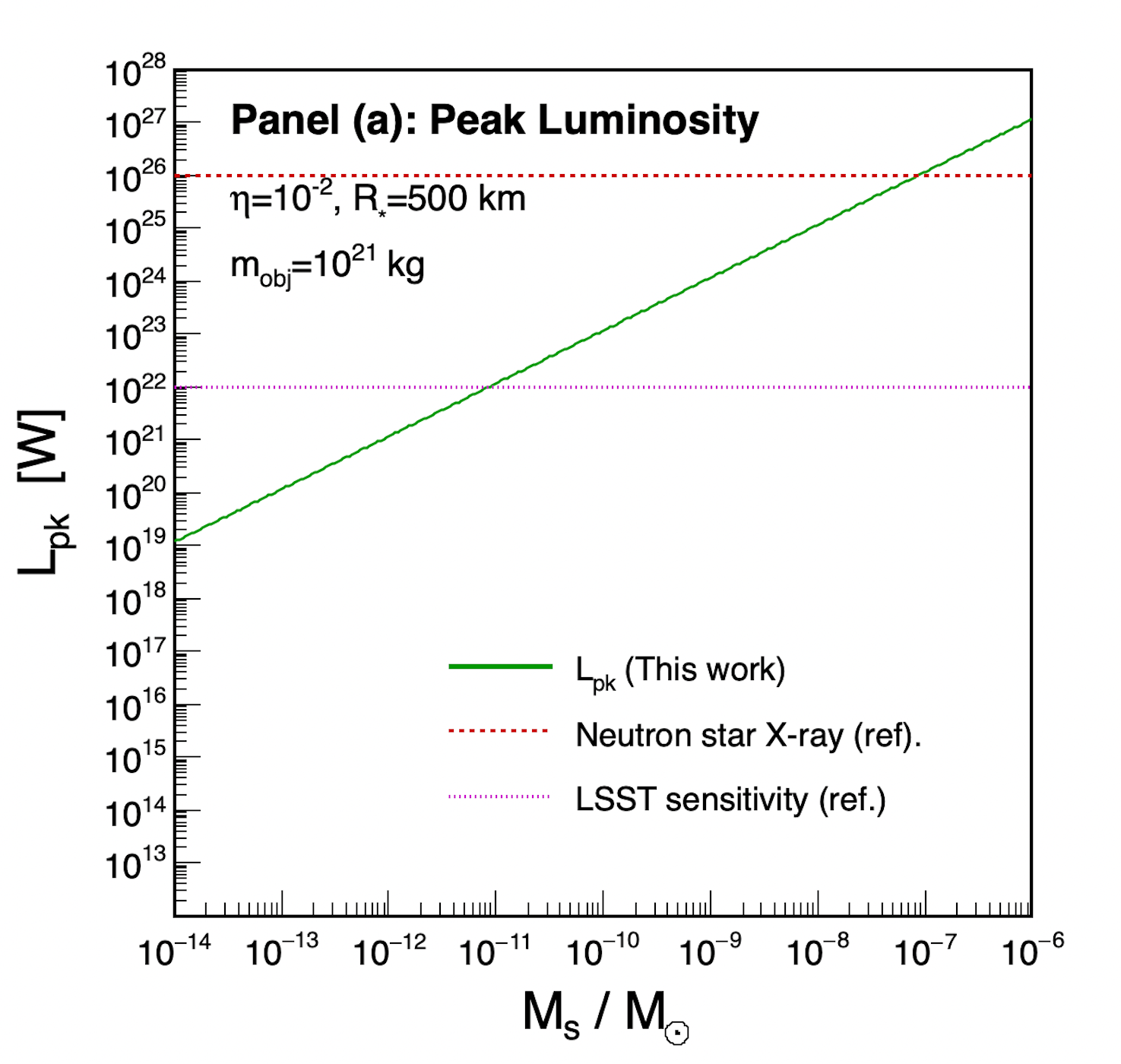}
    \hfill
    \includegraphics[width=0.48\textwidth]{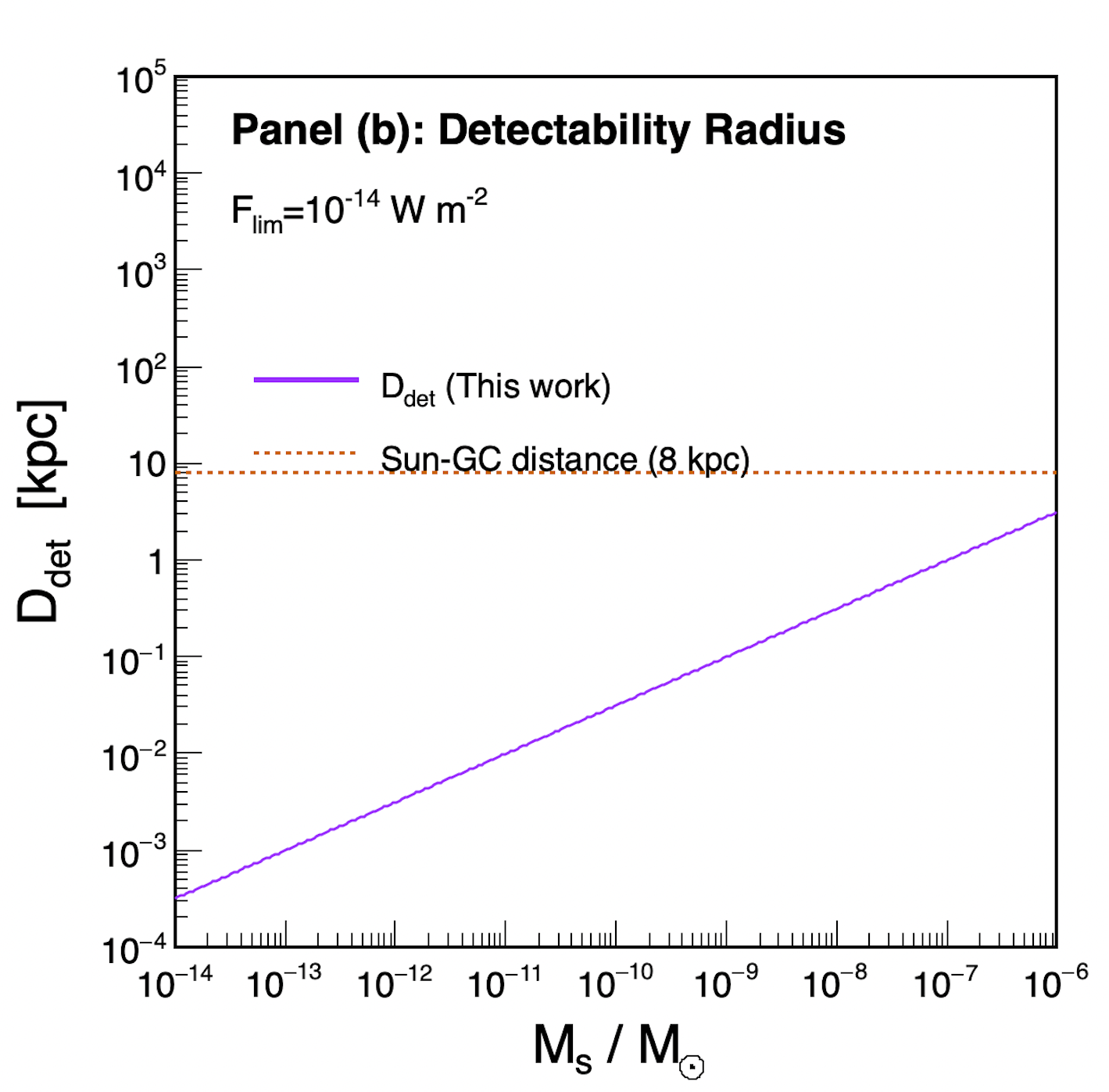}
    \caption{Flare luminosity and detectability for compact scalar clumps
    interacting with Kuiper Belt Objects. On the left we have the peak luminosity
    $L_{\rm pk}$ as a function of clump mass $M_s$, evaluated for
    $\eta=10^{-2}$, $R_*=500\,\mathrm{km}$ and
    $m_{\rm obj}=10^{21}\,\mathrm{kg}$. The neutron star reference
    luminosity is shown for comparison. Any survey flux threshold shown
    on this panel must be converted to a luminosity at an explicitly
    stated source distance. On the right, we have detectability radius
    $D_{\rm det}$ obtained from Eq.~\eqref{eq:Ddet} for
    $F_{\rm lim}=10^{-14}\,\mathrm{W\,m^{-2}}$, while the horizontal line marks
    the Sun-Galactic center distance of $8\,\mathrm{kpc}$.}
    \label{fig:flare_summary}
\end{figure*}
Panel~(a) of Figure~\ref{fig:flare_summary} shows that $L_\mathrm{pk}$ scales linearly with clump mass, $L_\mathrm{pk} \propto M_s$, reflecting the proportionality of the available gravitational interaction energy (Eq.~\ref{eq:Eavail}) to the clump mass at fixed KBO parameters. For the fiducial crossing timescale
$\tau = R_*/v \approx 2.3\times10^{3}\,\mathrm{s}$, a clump of mass $M_s = 10^{-10}\,M_\odot$ yields a peak luminosity of approximately $10^{22}\,\mathrm{W}$, consistent with the order of magnitude
estimate of Eq.~\ref{eq:Lpk}. This value coincides with the \textit{Rubin}/LSST single epoch sensitivity threshold      ~\cite{ivezic2019lsst}, establishing $M_s \simeq 10^{-10}\,M_\odot$ as the approximate lower mass boundary for electromagnetic detectability with current wide field time domain facilities. Clumps heavier than roughly $10^{-8}\,M_\odot$ produce flares that exceed the steady X-ray output of a young neutron star, placing them well within reach of existing X-ray and optical all sky monitors.
Together, panels~(a) and~(b) demonstrate that the ADAF flare channel
transitions from a purely local solar system probe to a genuinely
galactic scale dark matter detector at clump masses accessible to ultralight scalar field models~\cite{Hui:2017ltb, Liebling:2012fv}, making it a powerful and largely unexplored complement to microlensing ~\cite{niikura2017microlensing} and gravitational wave searches ~\cite{arvanitaki2020large} in this mass regime.

\section{ Atomic Clock Probe of Oscillating Scalar Dark Matter}
\label{sec:clocks}

Consider a light scalar field $\phi$ of mass \mphi\ that constitutes a
fraction \fphi\ of the local dark matter density \rhoDM. In the non-relativistic regime relevant for virialized galactic dark matter, the field oscillates approximately harmonically at an angular frequency equal to its mass,
\begin{equation}
  \phi(t) \simeq \phi_0\cos(\mphi\, t) \;,
\label{eq:phiosc}
\end{equation}
where $\phi_0$ is the (real) amplitude of the oscillation.
This form follows from the classical Klein-Gordon equation in a harmonic potential $V(\phi) = \tfrac{1}{2}m_\phi^2\phi^2$ when spatial gradients are negligible on the coherence scale set by the de~Broglie wavelength of the dark
matter~\cite{Hui:2017ltb,hees2016searching}.

The energy density stored in this coherently oscillating field can be written as
\begin{equation}
  \rho_\phi \simeq \tfrac{1}{2}m_\phi^2\phi_0^2 \;,
\label{eq:rhophi}
\end{equation}
obtained by averaging the sum of kinetic and potential contributions over many oscillation periods. For a given local dark matter density \rhoDM, the amplitude is fixed by the
requirement $\rho_\phi = f_\phi\rho_\mathrm{DM}$, which yields
\begin{equation}
  \phi_0 = \sqrt{\frac{2\rho_\phi}{m_\phi^2}}
          = \sqrt{\frac{2f_\phi\rho_\mathrm{DM}}{m_\phi^2}}
          = \frac{\sqrt{2f_\phi\rho_\mathrm{DM}}}{m_\phi} \;.
\label{eq:phi0}
\end{equation}
For estimates, we take
$\rho_\mathrm{DM} \approx 0.4\,\mathrm{GeV/cm^3}$, based on galactic
dynamics near the Solar neighborhood~\cite{Ferreira:2021}.
In the following, it is convenient to adopt the simplifying assumption
$f_\phi = 1$, so that the scalar field saturates the local dark matter
density. All bounds derived under this assumption can be trivially rescaled as $d_X \propto f_\phi^{-1/2}$ if $\phi$ constitutes only a subcomponent of dark matter~\cite{hees2016searching}.

Atomic and molecular transition frequencies depend on dimensionless
combinations of fundamental constants, such as the fine structure constant $\alpha$, the electron-to-proton mass ratio $\mu = m_e/m_p$, and ratios of light-quark masses to the QCD scale ~\cite{Kennedy:2020bac,Paolini:2018}. If the scalar field couples to these constants, its oscillations lead to fractional variations of the constants and, consequently, to time-dependent
shifts in clock frequencies~\cite{Kennedy:2020bac,hees2016searching}.

A generic way to parametrize such couplings is to write the fractional
variation of a fundamental constant $X$ as
\begin{equation}
  \frac{\delta X}{X} = d_X\,\frac{\phi}{\Lambda} \;,
\label{eq:dX}
\end{equation}
where $d_X$ is a dimensionless coupling constant and $\Lambda$ is a
high energy scale that characterizes the strength of the interaction (for example, a cutoff scale, mediator mass, or symmetry breaking
scale)~\cite{Kennedy:2020bac}. Substituting the oscillatory solution in Eq.~(\ref{eq:phiosc}) into Eq.~(\ref{eq:dX}) yields
\begin{equation}
  \frac{\delta X(t)}{X} = d_X\,\frac{\phi_0}{\Lambda}\cos(\mphi t)
  = d_X\,\frac{\sqrt{2f_\phi\rho_\mathrm{DM}}}{m_\phi\Lambda}\cos(\mphi t)\;.
\label{eq:deltaX}
\end{equation}
Thus each constant $X$ acquires a coherent sinusoidal modulation at angular frequency \mphi\ with amplitude proportional to
$d_X/(m_\phi\Lambda)$~\cite{Kennedy:2020bac,hees2016searching}.

The frequency $\nu_i$ of a clock $i$ depends on the set of constants $X$
through atomic structure. For sufficiently small variations, the fractional frequency shift can be written as a linear combination
\begin{equation}
  \frac{\delta\nu_i}{\nu_i} = \sum_X K^i_X\,\frac{\delta X}{X} \;,
\label{eq:dnu}
\end{equation}
where the sensitivity coefficients $K^i_X$ quantify how strongly the
transition responds to each constant~\cite{Paolini:2018,Wcislo:2018}.
These coefficients can be computed using atomic structure calculations and are tabulated for many commonly used optical and microwave clock transitions~\cite{Delva:2023, Wcislo:2018}.
State-of-the-art optical lattice clocks and single-ion clocks now reach fractional stabilities and accuracies at this $10^{-18}$ level, making them exquisitely sensitive probes of tiny oscillations in fundamental constants~\cite{Bloom:2014, Bothwell:2019, Roberts:2017}.

Clock experiments typically monitor the ratio of frequencies of two clocks, $R_{ij} = \nu_i/\nu_j$, which removes many common-mode systematic effects. The fractional modulation of the clocks can be written as
\begin{equation}
  \frac{\delta(\nu_i/\nu_j)}{\nu_i/\nu_j}
  = \sum_X\!\left(K^i_X - K^j_X\right)\frac{\delta X}{X} \;.
\label{eq:dratio}
\end{equation}
Inserting Eq.~(\ref{eq:deltaX}) into Eq.~(\ref{eq:dratio}) gives
\begin{equation}
  \frac{\delta(\nu_i/\nu_j)}{\nu_i/\nu_j}(t)
  = \sum_X\!\left(K^i_X - K^j_X\right)d_X
    \frac{\sqrt{2f_\phi\rho_\mathrm{DM}}}{m_\phi\Lambda}\cos(\mphi t)\;.
\label{eq:signal}
\end{equation}
The effect of the scalar dark matter field is therefore a coherent,
sinusoidal modulation of the clock-frequency ratio at angular frequency \mphi, with amplitude scaling as $\sqrt{f_\phi\rho_\mathrm{DM}}/(m_\phi\Lambda)$, as anticipated in the
heuristic argument~\cite{hees2016searching,Wcislo:2018}.

It is convenient to define, for each clock pair $(i,j)$, an effective
coupling that packages the dependence on individual constants:
\begin{equation}
  d^{(ij)}_\mathrm{eff} \equiv \sum_X\!\left(K^i_X - K^j_X\right)d_X \;.
\label{eq:deff}
\end{equation}
Then Eq.~(\ref{eq:signal}) simplifies to
\begin{equation}
  \frac{\delta(\nu_i/\nu_j)}{\nu_i/\nu_j}(t)
  = d^{(ij)}_\mathrm{eff}\,
    \frac{\sqrt{2f_\phi\rho_\mathrm{DM}}}{m_\phi\Lambda}\cos(\mphi t)\;.
\label{eq:signaleff}
\end{equation}
This expression directly relates the properties of the scalar dark matter ($m_\phi$, $f_\phi$, $\rho_\mathrm{DM}$) and its couplings ($d_X$, $\Lambda$) to the observable time series signal in clock comparisons~\cite{Kennedy:2020bac,Wcislo:2018}.
\\
\\
A practical data analysis to search for oscillating scalar dark matter with atomic clocks, starting with raw clock data and ending with bounds on the modulation amplitude $A_{ij}$ and the couplings $d_X$. We follow a similar methodology used in recent global optical clock and clock comparison searches, adapted to the theoretical framework of
Refs.~\cite{Delva:2023, kobayashi2022search, Kennedy:2024, Wcislo:2018}. We measure the frequency ratio 
\begin{equation}
  R_{ij}(t_k) = \frac{\nu_i(t_k)}{\nu_j(t_k)}
\label{eq:Rij}
\end{equation}
at discrete times $t_k$ (clock cycles), where $i$ and $j$ label two clocks or oscillators with different sensitivities $K^i_X$ and $K^j_X$ to the fundamental constant are labelled. For example kobayashi et al.\ compare a $^{171}$Yb optical lattice clock to a $^{133}$Cs fountain over 298~days, providing a long dataset suitable for ultralight dark matter searches ~\cite{kobayashi2022search}. Similarly, Wci\'{s}{\l}o et al.\ use four optical clocks in a global network ~\cite{Wcislo:2018}. If access to real data is not available, one can generate a realistic synthetic dataset by combining:
\begin{itemize}
  \item a constant mean ratio $R_0$ (e.g., unity);
  \item a small sinusoidal ``signal'' consistent with Eq.~(\ref{eq:signaleff}),
    $\Delta R_\mathrm{sig}(t) = A_{ij}\cos(m_\phi t + \varphi)$;
  \item stochastic noise drawn from a Gaussian distribution with variance set by the clock instability (e.g., Allan deviation scaling)~\cite{Wcislo:2018}.
\end{itemize}

In this study, we can isolate small fractional variations because it is convenient to define the fractional deviation of the frequency ratio from its mean value
\begin{equation}
  \Delta_{ij}(t) \equiv
  \frac{(\nu_i/\nu_j)(t) - \langle\nu_i/\nu_j\rangle}{\langle\nu_i/\nu_j\rangle} \;,
\label{eq:Deltaij}
\end{equation}
where $\langle\nu_i/\nu_j\rangle$ denotes the time-averaged ratio over the dataset. This construction removes any static offset and expresses the signal in terms of dimensionless fractional deviations at the $10^{-16}$--$10^{-18}$ level typical of state-of-the-art clocks~\cite{Bloom:2014,Bothwell:2019}. For a given trial scalar mass \mphi, corresponding to an angular frequency $\omega_\phi = m_\phi$, the predicted dark matter signal in Eq.~(\ref{eq:signaleff}) has the form of a sinusoid with angular frequency \mphi. Accordingly, the measured fractional deviation $\Delta_{ij}(t)$ can be modeled as
\begin{equation}
  \Delta_{ij}(t) = A_c\cos(m_\phi t) + A_s\sin(m_\phi t) + \text{noise} \;,
\label{eq:fitmodel}
\end{equation}
where $A_c$ and $A_s$ are the cosine and sine quadrature amplitudes, and the residual term encompasses statistical noise and any unmodelled systematics. For each chosen \mphi, Eq.~(\ref{eq:fitmodel}) defines a linear least squares fit problem for $A_c$ and $A_s$ using the time series data~\cite{kobayashi2022search,Kennedy:2024,Wcislo:2018}.

The total modulation amplitude associated with this sinusoid can be written as
\begin{equation}
  A_{ij} \equiv \sqrt{A_c^2 + A_s^2} \;,
\label{eq:Aij}
\end{equation}
which is directly comparable to the theoretical prediction in Eq.~(\ref{eq:signaleff}). Combining Eqs.~(\ref{eq:signaleff}) and (\ref{eq:fitmodel}), one finds that, if the observed modulation is entirely due to the scalar dark matter signal, the amplitude must satisfy
\begin{equation}
  A_{ij} = d^{(ij)}_\mathrm{eff}\,
            \frac{\sqrt{2f_\phi\rho_\mathrm{DM}}}{m_\phi\Lambda} \;.
\label{eq:Aij_theory}
\end{equation}
Equation~(\ref{eq:Aij_theory}) provides the key link between the
experimentally extracted amplitude $A_{ij}$ and the underlying properties of the scalar dark matter ($m_\phi$, $f_\phi$, $\rho_\mathrm{DM}$) and its couplings ($d_X$, $\Lambda$).

A search for an oscillatory signal at frequency \mphi\ proceeds by scanning over a grid of trial masses and, at each point, fitting the model in Eq.~(\ref{eq:fitmodel}) to obtain $A_c(m_\phi)$ and $A_s(m_\phi)$, and hence, $A_{ij}(m_\phi)$ via Eq.~(\ref{eq:Aij}).
In the absence of a statistically significant excess above the expected noise level, the result is an upper bound on the modulation amplitude as a function of mass,
\begin{equation}
  A_{ij}(m_\phi) \leq A^{\max}_{ij}(m_\phi) \;,
\end{equation}
at a chosen confidence level. Combining this empirical bound with the theoretical relation Eq.~(\ref{eq:Aij_theory}) yields an upper limit on the effective coupling.
\begin{equation}
  d^{(ij)}_\mathrm{eff}(m_\phi) \leq
  \frac{A^{\max}_{ij}(m_\phi)\,m_\phi\Lambda}{\sqrt{2f_\phi\rho_\mathrm{DM}}} \;.
\label{eq:deff_limit}
\end{equation}

If a single fundamental constant $X$ dominates the sensitivity of the clock pair, one may approximate $d^{(ij)}_\mathrm{eff} \simeq (K^i_X - K^j_X)\,d_X$. In that case, Eq.~(\ref{eq:deff_limit}) can be recast as a direct limit on the underlying coupling $d_X$,
\begin{equation}
  |d_X(m_\phi)| \leq
  \frac{A^{\max}_{ij}(m_\phi)\,m_\phi\Lambda}
       {\sqrt{2f_\phi\rho_\mathrm{DM}}\,|K^i_X - K^j_X|} \;.
\label{eq:dX_limit}
\end{equation}
Equations~(\ref{eq:deff_limit}) and~(\ref{eq:dX_limit}) provide the final mapping from experimental upper limits on clock-frequency modulation amplitudes to constraints on the couplings of oscillating scalar dark matter to fundamental constants. Existing and future clock comparisons over a wide range of modulation periods, from seconds to years, can therefore be straightforwardly interpreted as
bounds in the $(m_\phi, d_X)$ parameter space ~\cite{kobayashi2022search,Kennedy:2020bac,hees2016searching,Wcislo:2018}.
\begin{figure*}[t]
    \centering
    \includegraphics[width=0.8\linewidth]{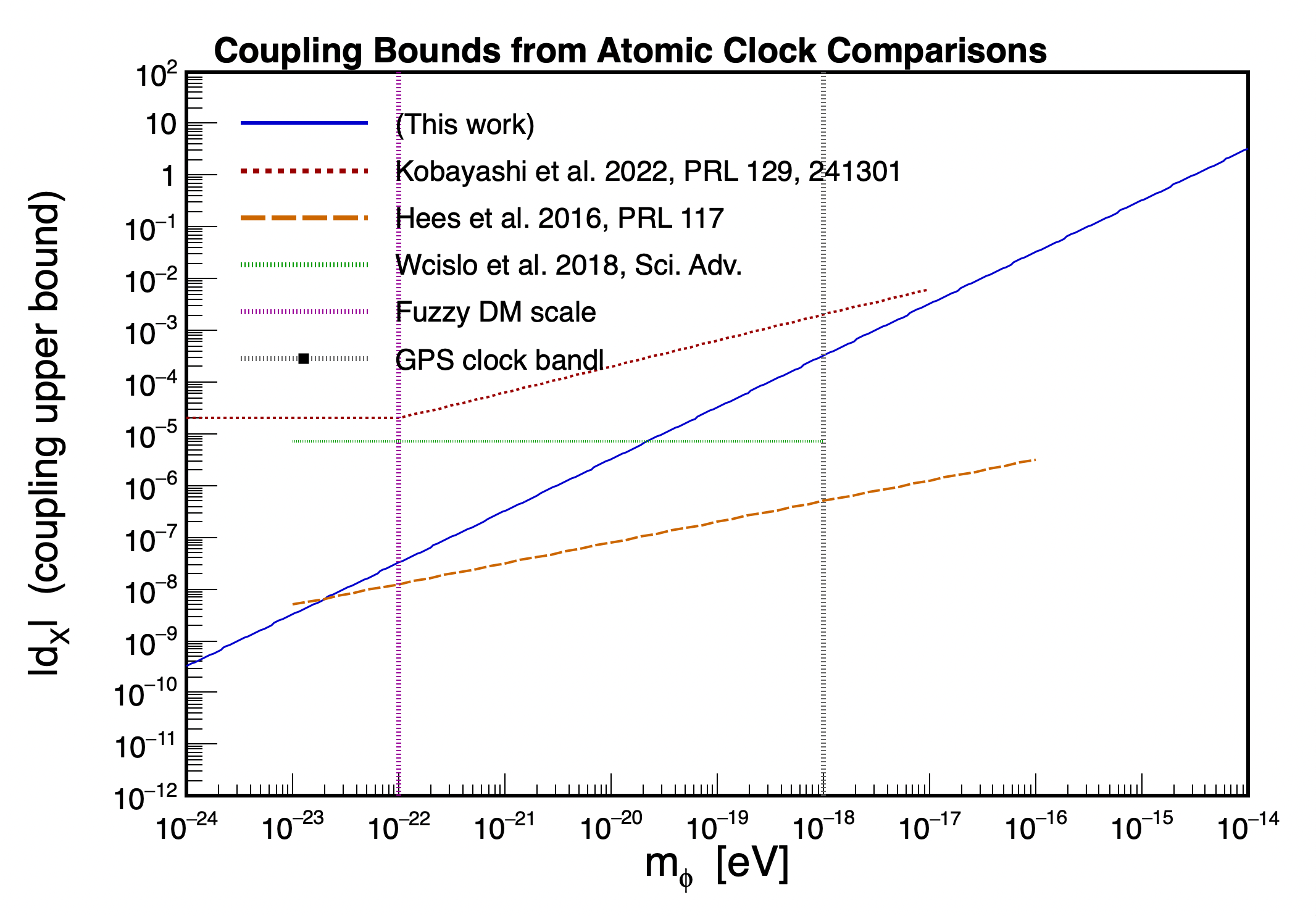}
    \caption{Upper bounds on the scalar dark matter coupling $|d_X|$ as a function of scalar mass $m_\phi$, derived from a null detection in atomic clock comparisons \ref{eq:dX_limit}. The solid blue curve is the projected sensitivity of a single clock-pair comparison with fractional instability $\sigma_{\rm clock} = 10^{-18}$ \cite{Bloom:2014, Bothwell:2019}, assuming the field saturates the local dark matter density ($f_\phi = 1$, $\rho_{\rm DM} = 0.4~\mathrm{GeV\,cm^{-3}}$).} 
    \label{fig:atomic_clock}
\end{figure*}
\section{Astrometric Lensing by Scalar Dark Matter Clumps}
The previous probes which we have considered so far were either luminous transients from compact scalar clumps or oscillatory variations of fundamental constants induced by a coherently oscillating scalar field. Now another possibility that we will consider here is that scalar dark matter forms sufficiently compact gravitating clumps, such as boson stars, solitonic cores or other localized scalar configurations and that these can act as dark astrometric lenses. What is interesting to note is that in this channel no direct nongravitational coupling to Standard Model fields is required and the observable is instead the apparent displacement of a background star as the scalar clump passes close to the line of sight.
\\
\\
Let the scalar dark matter be partitioned into compact clumps of mass $M_s$, and let these clumps make up a fraction $f_s$ of the local dark matter density $\rho_{\rm DM}$. If the clump mass function is taken to be approximately monochromatic, the number density of clumps is
\begin{equation}
n_s=\frac{f_s\rho_{\rm DM}}{M_s}
\end{equation}
This relation is the basic connection between the dark matter model and the lensing event rate, and the reason for that is that for fixed $f_s\rho_{\rm DM}$, heavier clumps are individually stronger lenses, but they are also rarer. This competition is what controls the final mass dependence of the astrometric constraint. We now consider a clump of mass $M_s$ at distance $D_L$ lensing a background source at distance $D_S$ and with that in mind, we define
\begin{equation}
D_{LS}=D_S-D_L
\end{equation}
For a compact point-like lens, the angular Einstein radius is
\begin{equation}
\theta_E= \left( \frac{4GM_s}{c^2} \frac{D_{LS}}{D_LD_S} \right)^{1/2}
\end{equation}
The corresponding physical Einstein radius in the lens plane is then given by 
\begin{equation}
R_E=D_L\theta_E = \left( \frac{4GM_s}{c^2}
\frac{D_LD_{LS}}{D_S} \right)^{1/2}
\end{equation}
Note that the angular scale $\theta_E$ grows with the square root of the clump mass and for nearby lenses it is enhanced because $\theta_E\propto D_L^{-1/2}$ when $D_L\ll D_S$. This is of interest because this is why local scalar clumps, which if present, can in principle produce larger angular deflections than distant clumps of the same mass. In the solar system or local halo limit where the lens is much closer to the observer than the source, we have $D_L\ll D_S$ and hence $D_{LS}/D_S\simeq 1$. So in this case, the Einstein angle comes to be
\begin{equation}
\theta_E\simeq \left( \frac{4GM_s}{c^2D_L} \right)^{1/2}
\end{equation}
while the physical Einstein radius becomes
\begin{equation}
R_E\simeq \left( \frac{4GM_sD_L}{c^2} \right)^{1/2}
\end{equation}
These local expressions show that a nearby clump has a large angular signature, but the corresponding path length through the local volume is small and this creates an important tradeoff between individual detectability and the probability of seeing an event. To see this clearly, let $\beta$ denote the angular separation between the unlensed source position and the lens. It is convenient to define the dimensionless impact parameter $u=\frac{\beta}{\theta_E}$ and then, for an unresolved pair of lensed images the centroid shift relative to the unlensed source position is given by
\begin{equation}
\delta\theta(u)=\frac{u}{u^2+2}\theta_E
\end{equation}
This is the main astrometric observable and lets be clear about what it achieves. Unlike photometric microlensing, which relies on a measurable change in flux, astrometric microlensing can remain sensitive even when the magnification is small and the signal appears as a time-dependent shift in the apparent position of a background star. The maximum centroid shift follows by differentiating $\delta\theta(u)$ with respect to $u$. It is easy to check that the maximum occurs at $u=\sqrt{2}$ and the corresponding maximum shift is $\delta\theta_{\rm max}=\frac{\theta_E}{\sqrt{8}}$. This gives a simple necessary condition for detectability and that is, that if the largest possible centroid shift is below the astrometric threshold, then no impact parameter can produce a detectable event. Thus for an effective astrometric threshold $\sigma_{\rm th}$, we would require \begin{equation}
\frac{\theta_E}{\sqrt{8}}\gtrsim \sigma_{\rm th}
\end{equation}
Here we should note that $\sigma_{\rm th}$ may also be taken as a multiple of the single epoch astrometric uncertainty. For example $\sigma_{\rm th}=q\sigma_{\rm ast}$ with $q\sim 3$ or $5$ or perhaps also as an effective threshold after combining multiple epochs and accounting for the scanning law. This is the condition which will end up determining the minimum lens mass that can be probed by astrometric centroid shifts. Substituting the expression for $\theta_E$ into the detection condition gives us the following useful inequality \begin{equation}
\frac{1}{8} \frac{4GM_s}{c^2} \frac{D_{LS}}{D_LD_S}
\gtrsim \sigma_{\rm th}^2
\end{equation}
Solving for $M_s$, one could obtain in a very straightforward way
\begin{equation}
M_s\gtrsim M_{\rm min}= \frac{2c^2\sigma_{\rm th}^2}{G} \frac{D_LD_S}{D_{LS}}
\end{equation}
This expression gives the minimum mass scale for a compact clump to generate a resolvable astrometric shift at a given lens geometry. The dependence on $\sigma_{\rm th}^2$ is very important because improving astrometric precision by one order of magnitude, would end up lowering the detectable mass threshold by two orders of magnitude. For a representative Galactic lensing geometry with the lens roughly halfway to the source, $D_L=D_{LS}=D_S/2$, the minimum mass would then become
\begin{equation}
M_{\rm min} \simeq \frac{2c^2\sigma_{\rm th}^2D_S}{G}
\end{equation}
Numerically, this would mean
\begin{equation}
M_{\rm min} \simeq 9.8\times 10^{-3}M_\odot
\left( \frac{\sigma_{\rm th}}{100 \,\mu{\rm as}}
\right)^2 \left( \frac{D_S}{1 \,{\rm kpc}}
\right)
\end{equation}
This estimate tells us that Gaia-like astrometry with an effective threshold of order $100,\mu{\rm as}$ is naturally sensitive to compact clumps with masses around $10^{-2}M_\odot$ and above for kiloparsec-scale source distances. It also tells us that better astrometry rapidly moves the threshold down to planetary and asteroid-like masses. For a nearby Solar System lens with $D_L\ll D_S$, the minimum mass would instead be $M_{\rm min}^{\rm local}
\simeq \frac{2c^2\sigma_{\rm th}^2D_L}{G}$ and a useful numerical form is
\begin{equation}
M_{\rm min}^{\rm local} \simeq 4.8\times10^{-9}M_\odot
\left(\frac{\sigma_{\rm th}}{100 \,\mu{\rm as}}
\right)^2 \left( \frac{D_L}{100 \,{\rm AU}}
\right)
\end{equation}
This tells us that nearby clumps can be detectable at much lower masses than distant Galactic clumps and to see that consider if $\sigma_{\rm th}=10,\mu{\rm as}$ and $D_L=100,{\rm AU}$, the detectable mass scale falls to approximately $5\times10^{-11}M_\odot$. The drawback of this logic is that the available local volume is extremely small compared with a kiloparsec scale Galactic line of sight. 
\\
\\
To move forward, what we would need is an estimate for the maximum detectable impact parameter and for that we note that detection requires $\delta\theta(u)\geq \sigma_{\rm th}$ and using the centroid shift expression, this becomes
\begin{equation}
\frac{u}{u^2+2}\theta_E\geq \sigma_{\rm th}
\end{equation}
For the ease of calculations, we also define \begin{equation}
y=\frac{\sigma_{\rm th}}{\theta_E}
\end{equation}
The threshold equation is then given by
\begin{equation}
\frac{u}{u^2+2}=y
\end{equation}
Rearranging this will give us a quadratic equation from which we can write
\begin{equation}
u_\pm=
\frac{1\pm\sqrt{1-8y^2}}{2y}
\end{equation}
A real solution for this would exist only for $y\leq \frac{1}{\sqrt{8}}$ and that is equivalent to the earlier condition $\theta_E/\sqrt{8}\geq\sigma_{\rm th}$. Note also that the larger root in this case will give us the outer boundary of the detectable astrometric tube around the line of sight. Thus the maximum dimensionless detectable impact parameter is given by \begin{equation}
u_{\rm max}= \frac{ 1+\sqrt{1-8(\sigma_{\rm th}/\theta_E)^2}}{2(\sigma_{\rm th}/\theta_E)
}
\end{equation}
The physical maximum impact parameter in the lens plane is then given by
\begin{equation}
b_{\rm max}=D_L\theta_Eu_{\rm max}
\end{equation}
This is the quantity which would define the transverse width of the detectable lensing tube. This means that a clump passing within $b_{\rm max}$ of the line of sight produces a centroid shift above the adopted astrometric threshold at some point during the event. In the case of a strong detectability limit $\theta_E\gg\sigma_{\rm th}$, one has $y\ll 1$ and the larger root simplifies to $u_{\rm max}\simeq \frac{\theta_E}{\sigma_{\rm th}}$, which would mean $b_{\rm max}\simeq D_L\frac{\theta_E^2}{\sigma_{\rm th}}$. Keeping all this in mind, we are led to \begin{equation} \label{bmax}
b_{\rm max} \simeq \frac{4GM_s}{c^2\sigma_{\rm th}}
\frac{D_{LS}}{D_S}
\end{equation}
This is a particularly useful asymptotic result for us and it is so because in the weak astrometric tail, the detectable physical impact parameter scales linearly with the clump mass and inversely with the astrometric threshold. The factor $D_{LS}/D_S$ accounts for the usual lensing geometry and tends to unity for nearby lenses. For a local lens with $D_L\ll D_S$, this becomes
\begin{equation}
b_{\rm max}^{\rm local}
\simeq \frac{4GM_s}{c^2\sigma_{\rm th}}
\end{equation}
The corresponding characteristic event duration is $t_{\rm ast}\sim \frac{2b_{\rm max}}{v_\perp}$ where $v_\perp$ is the transverse velocity of the lens relative to the line of sight. This timescale is important because a signal with duration shorter than the survey cadence will be strongly suppressed by the detection efficiency. The differential event rate per monitored source can be obtained by considering the rate at which clumps sweep across the lensing tube and so, for a line of sight element $dD_L$, the event rate would be given by
\begin{equation}
d\Gamma_\star= 2b_{\rm max}(D_L,M_s)v_\perp n_s(D_L)dD_L
\end{equation}
Using $n_s=f_s\rho_{\rm DM}/M_s$, the total event rate per monitored star becomes
\begin{equation}
\Gamma_\star(M_s)= \int_0^{D_S} 2b_{\rm max}(D_L,M_s)v_\perp(D_L) \frac{f_s\rho_{\rm DM}(D_L)}{M_s} dD_L
\end{equation}
This is the appropriate event rate analogue of the astrometric optical depth, and we should note here that the commonly written quantity involving $\pi b_{\rm max}^2$ is more naturally interpreted as an instantaneous probability while the event rate for a time domain survey would be controlled by the swept length $2b_{\rm max}v_\perp$. For a survey monitoring $N_\star$ stars for a time $T$ with detection efficiency $\epsilon_{\rm det}$, the expected number of detected astrometric events is
\begin{equation}
N_{\rm ev}= N_\star T\epsilon_{\rm det}\Gamma_\star
\end{equation}
From this, we can then write
\begin{equation}
N_{\rm ev} = N_\star T\epsilon_{\rm det}
\int_0^{D_S} 2b_{\rm max}(D_L,M_s)v_\perp(D_L)
\frac{f_s\rho_{\rm DM}(D_L)}{M_s} dD_L
\end{equation}
If one assumes constant $\rho_{\rm DM}$ and constant $v_\perp$, then this would become
\begin{equation}
N_{\rm ev}= N_\star T\epsilon_{\rm det} \frac{2f_s\rho_{\rm DM}v_\perp}{M_s} \int_0^{D_S} b_{\rm max}(D_L,M_s)dD_L
\end{equation}
A null detection in this case then gives us a Poisson upper bound and at approximately $95\%$ confidence, we may take $N_{\rm ev}\lesssim 3$ and then solving for $f_s$ would end up giving us
\begin{equation}
f_s(M_s) \lesssim \frac{3M_s}{ 2N_\star T\epsilon_{\rm det}\rho_{\rm DM}v_\perp \int_0^{D_S}b_{\rm max}(D_L,M_s)dD_L }
\end{equation}
This would be our central constraint formula as it maps a null astrometric microlensing search into an upper bound on the fraction of dark matter contained in compact scalar clumps of mass $M_s$. The dependence on $M_s$ enters both explicitly and through the maximum impact parameter $b_{\rm max}$ and then in the strong detectability regime, using \eqref{bmax} the line of sight integral becomes
\begin{equation}
\int_0^{D_S}b_{\rm max}dD_L= \frac{4GM_s}{c^2\sigma_{\rm th}} \int_0^{D_S} \left(
1-\frac{D_L}{D_S} \right)dD_L
\end{equation}
It is easy to then check that \begin{equation}
\int_0^{D_S}b_{\rm max}dD_L = \frac{2GM_sD_S}{c^2\sigma_{\rm th}}
\end{equation}
Substituting this into the null detection bound ends up leading us to
\begin{equation}
f_s \lesssim \frac{3c^2\sigma_{\rm th}}
{4GN_\star T\epsilon_{\rm det}\rho_{\rm DM}v_\perp D_S }
\end{equation}
In this asymptotic limit, the dependence on $M_s$ cancels and this happens because the number density decreases as $M_s^{-1}$ while the detectable impact parameter increases as $M_s$. Once we see that the clump is massive enough to lie deep in the astrometric tail regime, the survey sensitivity becomes approximately flat in $M_s$ too. For making numerical estimates we adopt $\rho_{\rm DM}=0.4\,{\rm GeV\,cm^{-3}}$, $v_\perp=200\,{\rm km\,s^{-1}}$,$D_S=1\,{\rm kpc}$, $N_\star=10^8$, $T=5,{\rm yr}$ and we set $\epsilon_{\rm det}=1$. With these choices, the high mass asymptotic bound then becomes
\begin{multline}
f_s \lesssim 7.1\times10^{-3} \left( \frac{\sigma_{\rm th}}{100 \,\mu{\rm as}}
\right)\left( \frac{N_\star}{10^8}
\right)^{-1} \left( \frac{T}{5\,{\rm yr}}
\right)^{-1} \left( \frac{\epsilon_{\rm det}}{1}
\right)^{-1} \\ \left(\frac{\rho_{\rm DM}}{0.4 \,{\rm GeV \,cm^{-3}}} \right)^{-1}
\left( \frac{v_\perp}{200 \,{\rm km \, s^{-1}}}
\right)^{-1} \left( \frac{D_S}{1 \, {\rm kpc}}
\right)^{-1}.
\end{multline}
This estimate tells us very interestingly that a Gaia-like astrometric survey can constrain compact scalar clumps at the percent level once the clump mass is large enough for the astrometric signal to exceed the threshold over a sizeable impact parameter range. The bound will also improve linearly with astrometric precision, survey duration, number of monitored stars and line of sight distance. The detectable mass threshold is here controlled very strongly by the effective astrometric precision. Note that for Galactic scale lensing with a representative source distance $D_S=1\,{\rm kpc}$, the minimum mass required for the maximum centroid shift to exceed the threshold would approximately be
\begin{equation}
M_{\rm min} \simeq 9.8\times10^{-3}M_\odot
\left( \frac{\sigma_{\rm th}}{100\,\mu{\rm as}}
\right)^2 \end{equation}
Thus, $\sigma_{\rm th}=100 \,\mu{\rm as}$ gives $M_{\rm min}\sim10^{-2}M_\odot$, while $\sigma_{\rm th}=30 \,\mu{\rm as}$ and $10 \,\mu{\rm as}$ give $M_{\rm min}\sim9\times10^{-4}M_\odot$ and $M_{\rm min}\sim10^{-4}M_\odot$, respectively. This tells us that present and near future astrometry probes compact scalar clumps from roughly planetary to substellar scales, with the low mass reach improving quadratically as the astrometric threshold is lowered. A crude numerical scan using the exact threshold condition for $u_{\rm max}$ would give us the following representative null detection limits on the clump fraction $f_s$, where in each row the three entries correspond to $\sigma_{\rm th}=100\,\mu{\rm as}$, $30\,\mu{\rm as}$, and $10\,\mu{\rm as}$
\begin{multline}
f_s(10^{-8}M_\odot)\sim 1.4\times10^2,\ 4.2\times10^1\,\ 3.7,\qquad \\
f_s(10^{-6}M_\odot)\sim 3.7\times10^1,\ 1.1,\ 4.2\times10^{-2}
\end{multline}
These values show us that very low mass clumps are mostly unconstrained by current astrometry, although a $10\,\mu{\rm as}$ threshold begins to reach the few percent level near $10^{-6}M_\odot$, while for larger masses the constraints can become substantially stronger too. The $M_s\sim10^{-2}M_\odot$ regime is particularly close to the natural threshold for Gaia like astrometry, where percent level bounds become possible. At still higher masses, the limits approach the high mass asymptotic regime because the increase in lensing cross section is largely compensated by the decrease in the number density of clumps. For a purely local Solar System lensing search, the shorter lens distance greatly lowers the detectable mass scale as at $D_L=100\,{\rm AU}$, one finds $M_{\rm min}^{\rm local}\simeq4.8\times10^{-9}M_\odot(\sigma_{\rm th}/100\,\mu{\rm as})^2$, which becomes $M_{\rm min}^{\rm local}\sim5\times10^{-11}M_\odot$ for $\sigma_{\rm th}=10\,\mu{\rm as}$. However, the available local path length is tiny compared with a Galactic line of sight, so the corresponding constraint on the ambient Galactic scalar clump fraction is weak and so, taking an effective local depth $L=100\,{\rm AU}$, the asymptotic local estimate is
\begin{equation}
f_s^{\rm local} \sim 7\times10^2
\left( \frac{\sigma_{\rm th}}{10\,\mu{\rm as}}
\right) \left( \frac{N_\star}{10^8}
\right)^{-1} \left( \frac{T}{10\,{\rm yr}}
\right)^{-1} \left( \frac{100 \,{\rm AU}}{L} \right)
\end{equation}
This is not competitive as a bound on the smooth Galactic scalar clump population and so we can say that the local channel is better interpreted as a possible discovery channel for an unusually nearby clump, a locally enhanced clump population or a Solar System bound scalar structure, rather than as the strongest statistical constraint on the global dark matter fraction. So far, we have assumed that the scalar clump behaves as a compact point lens but if the clump has a finite radius $R_s$, as expected for boson stars or solitonic scalar configurations, the point lens approximation is valid only when the detectable impact parameter is larger than the physical size of the clump. For an extended spherically symmetric lens, the deflection at impact parameter $b$ depends on the projected mass enclosed within $b$
\begin{equation}
\alpha(b) \simeq \frac{4GM_{\rm 2D}(<b)}{c^2b}
\frac{D_{LS}}{D_S}
\end{equation}
Writing $M_{\rm 2D}(<b)=M_sF(b/R_s)$ with $F(x)\rightarrow 1$ for $x\gg 1$ and $F(x)<1$ for $x\lesssim 1$, the astrometric detection condition for us will then become
\begin{equation}
\frac{4GM_s}{c^2b} F(b/R_s) \frac{D_{LS}}{D_S}
\gtrsim \sigma_{\rm th}
\end{equation}
The point mass result is recovered when $b_{\rm max}\gg R_s$. Note also here that if instead $b_{\rm max}\lesssim R_s$, the enclosed projected mass is reduced and the astrometric deflection is suppressed and then a rough compactness requirement for the point lens constraint to apply is therefore
\begin{equation}
R_s \lesssim \frac{4GM_s}{c^2\sigma_{\rm th}}
\frac{D_{LS}}{D_S}
\end{equation}
This condition is crucial for scalar dark matter and we should now note why as well. The astrometric lensing bounds constrain compact scalar clumps and not a smooth coherent scalar field. This means that diffuse scalar configurations with radii larger than the relevant impact parameter produce weaker centroid shifts and require a dedicated extended lens analysis. 
\\
\\
The crude estimates therefore indicate that a Gaia-like astrometric search can reach
\begin{equation}
f_s\lesssim 10^{-2}
\end{equation}
for compact scalar clumps with
\begin{equation}
M_s\gtrsim 10^{-2}M_\odot
\end{equation}
while improved astrometry at the level of $10$--$30,\mu{\rm as}$ can plausibly reach
\begin{equation}
f_s\lesssim 10^{-3}-10^{-2}
\end{equation}
for masses extending down to
\begin{equation}
M_s\sim 10^{-4}M_\odot-10^{-3}M_\odot
\end{equation}
\begin{figure*}[!t]
    \centering
    \includegraphics[width=0.8\linewidth]{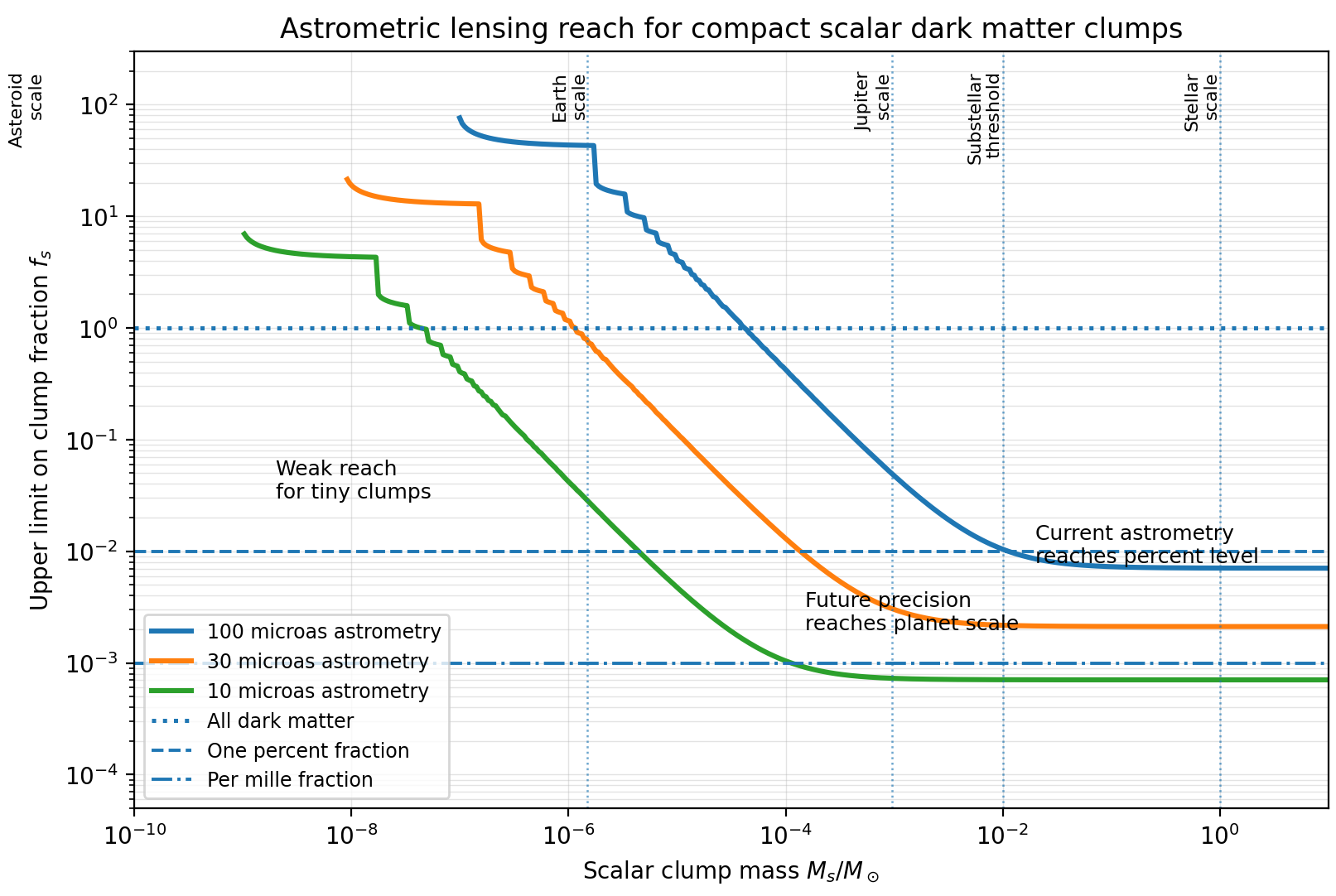}
    \caption{Projected astrometric lensing constraints on the compact scalar clump fraction $f_s$ as a function of clump mass $M_s$, showing the reach of current and future astrometric precision thresholds of $100\,\mu{\rm as}$, $30\,\mu{\rm as}$, and $10\,\mu{\rm as}$ for a fiducial survey with $N_\star=10^8$, $T=5\,{\rm yr}$, $D_S=1\,{\rm kpc}$, $v_\perp=200\,{\rm km\,s^{-1}}$, and $\rho_{\rm DM}=0.4\,{\rm GeV\,cm^{-3}}$.}
    \label{astrom}
\end{figure*}
In Fig. \ref{astrom} we have plotted the estimated upper limits on the compact scalar clump fraction $f_s$ as a function of the clump mass $M_s$ for representative astrometric sensitivities. This tells us very clearly that current Gaia like precision can begin to probe compact scalar clumps at the percent level for masses near $10^{-2}M_\odot$ and above, while future precision at the level of $10$ to $30\,\mu{\rm as}$ can extend the reach to lower mass clumps and stronger bounds. The flattening of the curves at high masses shows the regime in which the increase in lensing cross section is approximately compensated by the reduced number density of heavier clumps. The indicated mass scales also show us the implications for different compact dark matter candidates, ranging from asteroid scale objects to planetary, substellar and stellar mass scalar clumps.
\\

\section{Conclusions}
In this work, we have tried to explore a set of Solar System and local Galactic probes for scalar field dark matter, with particular emphasis on ultralight scalar candidates that may appear either as coherently oscillating fields or perhaps also as compact self gravitating clumps. We first considered ADAF-like luminosity transients which are triggered by encounters between compact scalar clumps and Kuiper Belt Objects, deriving order of magnitude estimates for the trigger rate, peak luminosity, detectability radius and the corresponding bound on the clump fraction $f_s(M_s)$. We then reviewed how oscillating scalar dark matter can induce coherent variations of fundamental constants and then ended up showing how atomic clock frequency ratio measurements can be mapped into constraints on effective scalar couplings. Finally, we developed the astrometric microlensing channel for compact scalar clumps, deriving the centroid shift threshold, detectable impact parameter, event rate and null detection constraints. For fiducial Galactic astrometric searches, we found that Gaia-like precision can reach $f_s\lesssim10^{-2}$ for compact clumps near $M_s\gtrsim10^{-2}M_\odot$, while future $10$-$30\,\mu{\rm as}$ astrometry can plausibly extend the reach toward $f_s\sim10^{-3}$--$10^{-2}$ for lower mass compact clumps.
\\

The broader significance of our results here is that Solar System and local precision probes can provide a complementary window into scalar dark matter models, especially in regimes not fully covered by conventional cosmological, direct detection or photometric microlensing searches. The three channels studied here are sensitive to different physical aspects of the scalar sector as well and that adds to the novelty. The flare searches probe compact, energetically interacting clumps while the atomic clocks would probe coherently oscillating field components through possible couplings to Standard Model parameters and astrometric lensing probes compact gravitational substructure independently of nongravitational couplings. This complementarity means that future null results can be translated into multi channel bounds on the abundance, compactness, coupling strength and mass scale of scalar dark matter while a candidate signal in one channel could be cross checked against the expected response in the others. In particular, future time domain surveys, improved clock networks and high precision astrometric missions could turn the solar system into a sort of useful laboratory for constraining high energy physics motivated dark matter models.
\begin{acknowledgments}
We gratefully acknowledge support from Vanderbilt University and the U.S. National Science Foundation. The work of OT is supported in part by the Vanderbilt Discovery Doctoral Fellowship. The work of AG is supported in part by NSF Award PHY-2411502. We thank Robert J. Scherrer for helpful discussions.
\end{acknowledgments}

\bibliographystyle{apsrev4-2}
\bibliography{refs}

\end{document}